\documentclass[a4paper,11pt]{article}
\pdfoutput=1 

\usepackage{jheppub}
\usepackage{microtype}   
\usepackage[utf8]{inputenc}
\usepackage{multirow, amsthm}
\usepackage{braket}
\usepackage{mathrsfs}
\usepackage{tikz}
\usepackage[caption=false]{subfig}
\usepackage{xspace}
\usepackage{xcolor}
\usepackage{float,color}
\usepackage[separate-uncertainty=true]{siunitx}
\usepackage[ruled,norelsize]{algorithm2e}
\usepackage[countmax]{subfloat}
\usepackage{accents}
\usepackage{physics}
\usepackage{appendix}
\usepackage[capitalise]{cleveref}    

\newcommand{\fastjet}[1]{\textsc{FastJet\xspace #1}}
\newcommand{\pythia}[1]{\textsc{Pythia\xspace #1}}

\title{Tag N' Train: A Technique to Train Improved Classifiers on Unlabeled Data}

\author[1]{Oz Amram}
\author[1]{Cristina Mantilla Suarez}

\affiliation[1]{Department of Physics and Astronomy, The Johns Hopkins University, Baltimore, MD 21218, USA}
\emailAdd{oamram1@jhu.edu}

\abstract{
There has been substantial progress in applying machine learning techniques to classification problems in collider and jet physics. 
But as these techniques grow in sophistication, they are becoming more sensitive to subtle features of jets that may not be well modeled in simulation. 
Therefore, relying on simulations for training will lead to sub-optimal performance in data, but the lack of true class labels makes it difficult to train on real data. 
To address this challenge we introduce a new approach, called Tag N’ Train (TNT), that can be applied to unlabeled data that has two distinct sub-objects.
The technique uses a weak classifier for one of the objects to tag signal-rich and background-rich samples.
These samples are then used to train a stronger classifier for the other object.
We demonstrate the power of this method by applying it to a dijet resonance search. 
By starting with autoencoders trained directly on data as the weak classifiers, we use TNT to train substantially improved classifiers. 
We show that Tag N' Train can be a powerful tool in model-agnostic searches and discuss other potential applications.
}

\begin{document}
\maketitle

\flushbottom

\section{Introduction}
\label{sec:intro}

Despite numerous searches for physics beyond the standard model~\cite{ATLAS:EXO, ATLAS:HDB,CMS:EXO, CMS:B2G, LHCb:EXO}, the experiments at the Large Hadron Collider have not yet provided evidence for new particles or new fundamental forces of nature.
While ATLAS, CMS and LHCb have explored many possible signatures, their searches have often been tailored with a particular signal model in mind and have left unexplored a variety of final states ~\cite{Rappoccio:2018qxp,Kim:2019rhy,Craig:2016rqv}.
Given that it is impractical to perform a dedicated search for every possible new physics signature, it is natural to consider designing model-agnostic searches that make minimal assumptions about the nature of the signal.
Such searches have traditionally been performed at collider experiments by comparing distributions in data to simulation in many different final states~\cite{Abbott_2000,Abbott_2001,Abazov_2001,Aaron:2008aa,Aktas_2004,Aaltonen_2008, Aaltonen_2009,CMS:2017yoc,CMS:2011fra,Aaboud:2018ufy,ATLAS:2014sxa,ATLAS:2012qna}.
However, this technique is insensitive to signals with very small cross sections or in final states not well modeled by simulation. 

At the same time, classifiers used to tag hadronic jets of different types have greatly increased in performance thanks to the use of machine learning~\cite{Larkoski:2017jix, Almeida:2015jua, Baldi:2016fql, Kasieczka:2017nvn, Louppe:2017ipp, Kasieczka:2019dbj, Butter:2017cot, Cogan:2014oua, deOliveira:2015xxd, Komiske:2016rsd, Macaluso:2018tck,Komiske:2018cqr}, but in almost all applications, classifiers have been trained using simulation. 

Because simulations do not perfectly model the actual radiation pattern in jets~\cite{Barnard:2016qma}, their use in training will lead to sub-optimal performance on real data.
Training a model directly on data seems to be the only way not to compromise performance, but this is challenging due to the lack of true labels.

These two predicaments naturally motivate further explorations of how to directly train classifiers on actual data, and how to use them to perform model-agnostic searches on LHC data.

In \cite{Metodiev:2017vrx}, the Classification Without Labels (CWoLa) approach was introduced as a way to train classifiers directly on data.
Instead of relying on fully labeled examples, the CWoLa approach trains by using statistical mixtures of events with different amounts of signal; allowing the use of many of the techniques from fully-supervised training.
To apply this technique in practice, one must find information orthogonal to the classification task that can be used to select the mixed samples in data.
One application of this technique has been in a model-agnostic dijet resonance search~\cite{Collins:2018epr,Collins:2019jip}.
In this approach, called CWoLa hunting, a particular invariant mass region is used to select the potentially signal-rich sample and neighboring sidebands are used to select the background-rich sample.
This method has been shown to be highly sensitive to resonant new physics, but it is unclear how to extend the approach to non-resonant signals where the anomaly is not localized in a particular kinematic distribution, such as the resonance mass.
The technique further relies on the information used in classification being completely uncorrelated with the resonance mass. 
Slight correlations may lead to the classifier preferentially selecting background events at a particular resonance mass as signal like, distorting the distribution of background events and complicating the background estimate.

Another approach that has been explored~\cite{Heimel:2018mkt, Farina:2018fyg,Blance:2019ibf, Roy:2019jae} is to scan for anomalous events by using autoencoders trained directly on data.
Autoencoders are a type of network that learns how to compress an object to a smaller latent representation and then decompress it to reconstruct the original object.
An autoencoder trained on a background-rich sample can learn how to compress and decompress objects in background events, but will not learn to do the same for anomalous events. 
The reconstruction loss, the difference between the original and reconstructed representation of the object, can then be used as a classification score that selects anomalous signals.
While the autoencoders have the advantage of making very minimal model assumptions about the signal, their  signal-vs-background classification performance is worse than a dedicated classifier. 
This is because their training aim is compression and decompression, not classification; they learn how to effectively represent the data's dominant component (i.e. background events), but they do not learn anything about what the sought-after signal looks like.

Recent proposals have also been made to use other machine learning techniques for anomaly searches with varying degrees of model independence \cite{ DAgnolo:2018cun,DAgnolo:2019vbw,Cerri:2018anq,Hajer:2018kqm,DeSimone:2018efk,Mullin:2019mmh,Casa:2018avf,Dillon:2019cqt,Aguilar-Saavedra:2017rzt, Nachman:2020lpy, Andreassen:2020nkr}. See \cite{Nachman:2020lpy} for an overview of recent techniques.

We propose a technique for training classifiers on data that utilizes the CWoLa paradigm to improve weak classifiers. 
The method, called Tag N' Train (TNT), is based on the assumption that signal events contain two separate objects; and thus the appearance of these objects is correlated. 
By using the weak classifier and one of the objects, one can \textit{tag} events as signal-like or background-like.
This then provides samples of signal-rich and background-rich events which can be used to \textit{train} a classifier for the other object. 
This technique has a natural application to a model-agnostic searches for new physics in di-object events. 
We explore a dijet resonance search based on TNT that uses autoencoders as the initial weak classifiers.
We find that its sensitivity compares favorably to that of CWoLa hunting and autoencoder-based approaches. 
We also highlight that the TNT approach naturally allows data-driven ways to estimate QCD backgrounds in anomaly searches. 

This paper is organized as follows. 
Section~\ref{sec:TNT} outlines the Tag N' Train technique and its key assumptions. 
The remainder of the paper illustrates the power of TNT through an example based on a dijet search using jet substructure.
Section~\ref{sec:methods} describes the simulation and deep learning setup. 
Section~\ref{sec:search} emulates an LHC dijet anomaly search and includes signal sensitivity comparisons of the TNT technique to the CWoLa hunting and autoencoder based searches. 
Conclusions and possible future applications of the TNT approach are discussed in~\ref{sec:conclude}.

\section{Tag N' Train}
\label{sec:TNT}

The Tag and Train technique is a method for training classifiers directly on data. 
The technique assumes that the data has 2 distinct objects and, each of the objects can be ``tagged'', i.e. each has a weak classifier can select signal-like events.
It takes as input a set of unlabeled data events and the initial classifiers, and outputs two new classifiers which may be substantially improved.
The original classifiers might be trained directly on data with an unsupervised approach (e.g. autoencoders), trained on simulation that is known to mis-model data, or might be single features of the data which are known \textit{a priori} to be useful in distinguishing signal vs. background.

The main idea behind the approach is to exploit the paired nature of the data, where one can use one sub-component of the data to \textit{tag} examples as signal-like or background-like. 
These signal-rich and background-rich samples can then be used to \textit{train} a classifier for the other sub-component.
The approach bears some similarity to the commonly used Tag and Probe technique, that uses the two body decays of resonances to measure efficiencies in data \cite{Sirunyan:2018fpa, Aad:2016jkr}. 

TNT needs a consistent decomposition of the data into two sub-components hereafter named Object-1 and Object-2. 
It assumes one has initial classifiers for Object-1 and Object-2. 
It is worth pointing out that the technique can work if the data has more than two sub-components by combining multiple sub-components into a single 'Object' for the purposes of classification\footnote{E.g. if the data  has natural components A,B, and C, one could take Object-1 to include both A and B. Then the classifier for Object-1 would receive as input the features of A and B and produce a single classification score.}.

\begin{figure}
    \centering
    \includegraphics[width = 0.8 \textwidth]{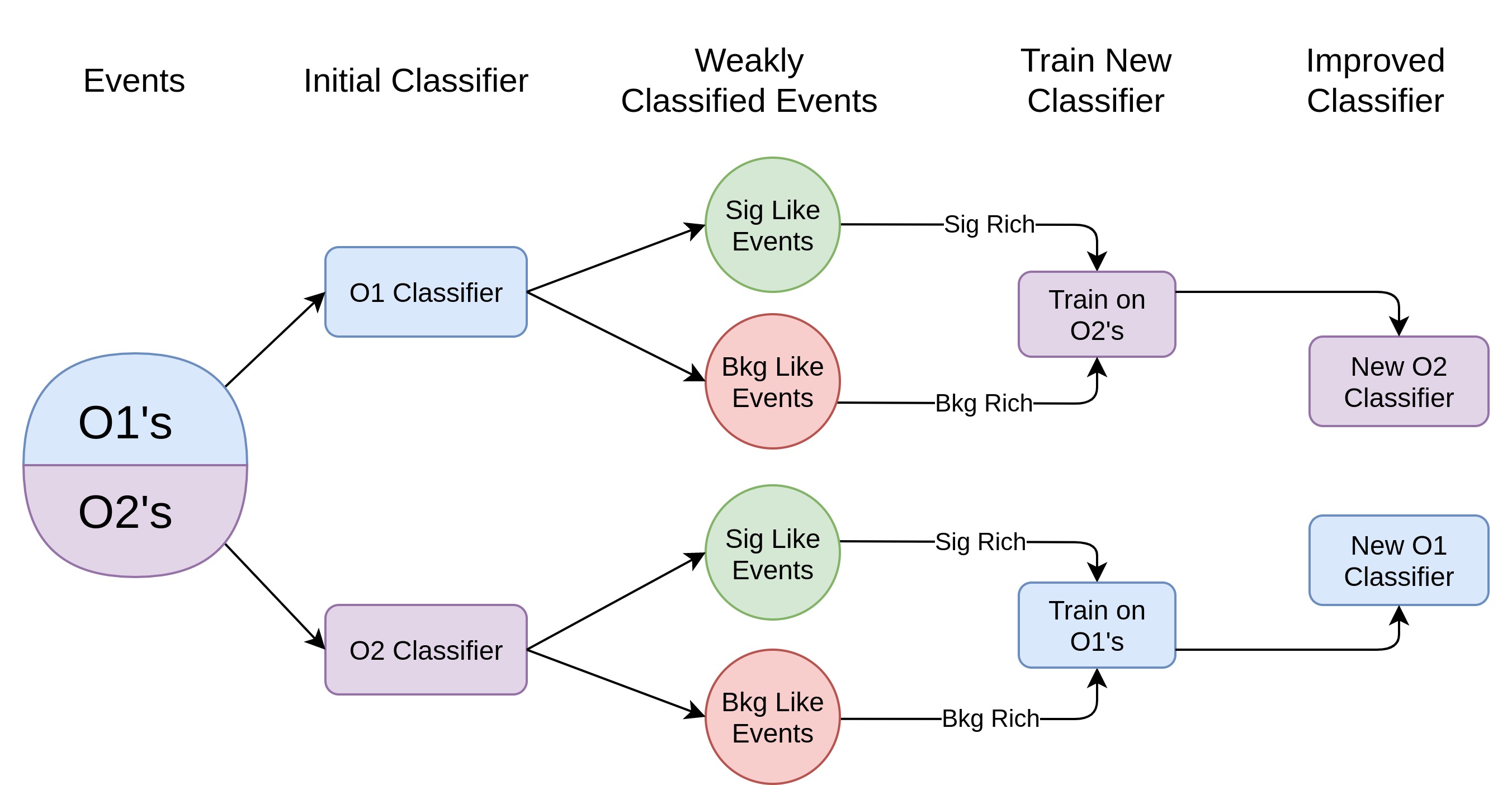}
    \caption{An illustration of the Tag N' Train technique. Here O1 and O2 represent Object-1 and Object-2, the two components of the data one wishes to train classifiers for. }
    \label{fig:TNT_algo}
\end{figure}

The procedure to train new classifiers is as follows: 

\begin{enumerate}
\item Classify events as signal-like or background-like using the Object-1's in each event.
\item Train a classifier to distinguish between the Object-2's in the signal-rich sample and the Object-2's in the background-rich sample.
\item Repeat the procedure, constructing samples by classifying the Object-2's in each event, and training a classifier for Object-1 using these samples~\footnote{One could also use the new classifier trained for Object-2 to train the classifier for Object-1, avoiding the need for initial classifiers for both objects. But this risks correlating the two classifiers.}.
\end{enumerate}
The Tag N' Train sequence is shown graphically in Figure~\ref{fig:TNT_algo}.

We stress that the signal-rich and background-rich samples obtained from the original classifiers do not need to be very pure for the technique to work.
In our example use case, detailed in Sec~\ref{sec:search}, we find that even for a few percent signal in the signal-rich sample, TNT can improve on the original classifiers.
This allows one to use weak classifiers as inputs, 
and/or apply the technique to data samples with very small amounts of signal present. 

One of the main assumptions behind this technique is that the information used for the classification tasks of Object-1's and Object-2's is uncorrelated in background events. 
This ensures the background objects in the signal-rich and background-rich samples have identical distributions even though they were selected using the other object in the event. 
If this is not the case, then the classifier may learn the difference between the background objects in these two samples, instead of learning information about the signal. 
If this condition is satisfied, then the requirements of the CWoLa paradigm are fulfilled and the classifier will be asymptotically optimal.
Additionally, the scores of the classifier for object 1 and object 2 on background events will remain uncorrelated, which is often desirable. 
In practice, one can afford to slightly violate this condition and still achieve good results, as long as the difference between the background in the two samples is smaller than the difference caused by the presence of signal. 

The technique works better if the initial classifier can create a larger separation between signal and background in the mixed samples used for training. 
Thus, if the TNT's output classifiers achieve better separation than the starting ones, multiple iterations of this technique can further improve classification performance until a plateau is reached.

In Appendix~\ref{sec:same_jet_test} we show why the Tag N' Train approach is appropriate, because it is not possible to use only one object to train an improved classifier on unlabeled data.

\section{Methods}
\label{sec:methods}

\subsection{Sample Generation}

To test our search strategy we use the research development dataset from the LHC Olympics 2020 challenge~\cite{Kasieczka_LHCO_RD}. 
The dataset consists of 1M QCD dijet events and 100k W' $\rightarrow$ XY events, both produced with \pythia 8~\cite{Sjostrand:2006za, Sjostrand:2007gs} with no pileup or multiple parton interactions included. 
The W' has a mass of 3.5 TeV, and the X and Y have masses of 500 GeV and 100 GeV respectively and both decay promptly to pairs of quarks.
Because of the large Lorentz boost, the hadronic decays of the X and Y bosons can each be captured in a single large radius jet.

Detector simulation is performed with Delphes 3.4.1~\cite{deFavereau:2013fsa} and particle flow objects are clustered into jets using the \fastjet~\cite{Cacciari:2005hq, Cacciari:2011ma} implementation of the anti-$k_t$ algorithm~\cite{Cacciari:2008gp} with a radius parameter of $R$ = 1.0.

For every event, we construct separate jet images~\cite{Cogan:2014oua,deOliveira:2015xxd,Komiske:2016rsd,Kasieczka:2017nvn,Macaluso:2018tck} for the two highest $p_T$ jets, to be used in event classification. 
Following~\cite{Macaluso:2018tck}, we apply pre-processing steps to our jets images before they are pixelated. 
We center the image based on the $p_T$ weighted center of the jet constituents and rotate the jet so that the principle axis is in the 12 o'clock position. 
Then the image is flipped along both axes so that the hardest $p_T$ constituent of the jet is in the upper left quadrant of the image.
After these steps, the image is pixelated into a 40 x 40 pixel image. 
The image covers an $\eta$ range and $\phi$ range of -0.7 to 0.7 around the center of the jet. 
In order to reduce dependence on the $p_T$ of the jet, each image is normalized so that the sum of all the pixel intensities sums 1. 
The sample of images is then normalized so that each pixel has zero mean and unit variance. 

\subsection{Architectures}
\label{sec:archs}

We use neural networks built and trained in Keras~\cite{chollet2015keras} with a TensorFlow~\cite{tensorflow2015-whitepaper} backend for all the classifiers considered in this work. 
All networks are trained with the Adam optimizer with a learning rate of 0.001, first and second moments decay rates of 0.8 and 0.99 respectively and a learning rate decay of 0.0005. 
Unless otherwise stated, all nodes use a Rectified Linear Unit (ReLu) activation function.

As this work is a proof of concept for the Tag N' Train technique, none of the network architectures have been optimized.

To train the autoencoders we use a convolutional network with filter sizes of 3x3 where the image's dimensionality is reduced through max pooling layers after each convolutional layer. 
The output is then fed through a dense layer which outputs the latent representation.
Based on~\cite{Farina:2018fyg, Heimel:2018mkt} we choose the size of our latent dimension to be 16 as this was seen to be within the performance plateau in both.
Then the architecture is mirrored, with 2D sampling layers in place of the max pooling layers to output an image of the same dimensions.  
We use a Mean Squared Error loss function during the classifier training. 

To train the image based classifiers we also use a convolution network with filter sizes of 3x3 for all convolutional layers followed by dense layers. 
The final layer has a sigmoid activation function. 
A binary cross-entropy loss is used during training. 

\section{A Dijet Anomaly Search}
\label{sec:search}

\subsection{Search Strategy}
One exciting application of the Tag N' Train technique is two-body searches at the LHC. 
We consider specifically a dijet anomaly search where one uses a autoencoder trained directly on data as the initial classifier, then uses the TNT technique to train improved classifiers. 
These improved classifiers are then used to suppress QCD backgrounds and a resonance is searched for in the invariant mass of the dijet events. 

We implement the search strategy as follows. 
For each event, we consider the highest two p$_T$ jets to be the dijet candidate.
In order to apply the Tag N' Train technique, we treat our Object-1 as the more massive jet in each event and Object-2 as the less massive jet. 
For each event we then have a separate image for the heavier and lighter jet.
To evaluate how well the strategy works with varying levels of signal, we vary the amount of signal present in the dataset by filtering out signal events. 
We run the search in the case where 9\%,1\% 0.3\% and 0.1\% of the events in the dataset are signal \footnote{These correspond to an S/B in the dijet mass range 3250 to 3750 GeV of 35\%, 6\%, 1.7\% and 0.5\% respectively)}.

Using an initial sample of 200k events we train separate autoencoders for the heavier and lighter jets in the sample. 
We use the autoencoder architecture described in section~\ref{sec:archs} for both autoencoders, use 10\% of events for validation, and train for 30 epochs.
We then use these autoencoders and a new sample of 200k events to train new classifiers with the Tag N' Train technique. 
Specifically, we define our signal-rich samples as the 20\% of events with the highest autoencoder loss, and the background-rich samples as the 40\% of events with the lowest autoencoder loss. 
We iterate the TNT procedure for a total of 3 iterations, each time using the classifiers from the previous iteration and a new set of 200,000 events as the inputs.
For the second iteration onward we use the 10\% events with the highest classifier score for our signal-rich sample, but still use the same selection for the background-rich sample.
We did not extensively optimize these selections, but did check that the performance of the technique is robust to the exact values used\footnote{For future work, it might be interesting to explore the use of a ``soft'' labeling scheme where the label for each event is not required to be strictly 0 or 1, but allowed to be a value in between based on the score of the initial classifier.}.

Because we are searching for a resonance we have additional information about the nature of our signal: it is likely to be localized to particular region in the dijet mass spectrum.
In such cases, where one has \textit{a priori} assumptions about the anomalous events, one can add them as additional selection requirements for events to be signal-like. 
Specifically, we require the events in the signal-rich sample to fall within a dijet invariant mass window in addition to the cut on the classifier score.
In a real search in data, one would scan the dijet mass range, training a separate set of networks for each dijet mass window (as is suggested in~\cite{Collins:2018epr,Collins:2019jip}), but we simplify things here by just requiring the dijet mass to be near the resonance mass of 3.5 TeV, i.e. within 3.3 and 3.7 TeV.
We do not apply any dijet mass selection to the background-rich sample, so that events from this sample populate this mass window as well. 
These additional requirements improve the fraction of signal events in the signal-rich sample. 
That is, for the dataset with 0.3\% signal events, there is ~1\% signal in the signal-rich samples without a dijet mass cut, ~1.5\% signal in the mass window with no additional requirement, and ~5\% when both are used together.

Because the TNT technique creates separate classifiers for each jet, one must combine them in a sensible way in order to select anomalous events.
Given the unsupervised nature of the search, one will not know \textit{a priori} what the optimal event selection will be for each object, and likely multiple criteria will be tried. 
We select events by choosing a certain percentile $X$ and requiring that the respective classifier score for both jets in that event to be above that percentile in their respective distributions. 
For example, if we pick a percentile selection of $X=20\%$ for our search, we select events where jet-1 must be in the top 20\% of jet-1 scores and jet-2 must be in the top 20\% of jet-2 scores.
One can scan over this percentile to produce a selection of the desired efficiency.
\footnote{This approach for combining the jet scores makes sense when both scores have roughly equal discrimination power and a ``tight'' selection is desired.
In other use cases one might want to pursue a different strategy.}

Once the classifiers have been trained, we select signal-like events from a sample of 300k events not used in training. 
To extract the final p-value of the anomaly search, we explore two different methods. 

The first is a count-and-count method, where one compares the number of selected events to an estimated number of background estimate. 
Because our signal events are selected using two independent classifiers, one can estimate the amount of background using an ABCD-like approach. 
This approach uses a jet-2 selection to obtain a sample of background-like events which are then used to measure the efficiency of the jet-1 selection on background, and vice versa. 
We select our background-like sample using the 40\% of scores with the lowest jet-2 scores and measure the rate of background events having jet-1 scores above the selection threshold.
The same procedure can then be used to measure the rate of background events passing the jet-2 selection.
Once the background efficiency of the jet-1 and jet-2 selection has been measured, the number of background events in the signal region can be estimated by multiplying these two efficiencies by the total number events before selection. 
A p-value can then be computed based on the number of observed events in the signal region and the number of estimated background events.

The second approach is a simple bump-hunt; applicable only to resonant signals.
After selecting the signal-like events we bin this dataset in $M_{jj}$ and perform the signal extraction.
We model the shape of the falling QCD background component with a third degree polynomial and propagate the uncertainties on the polynomial parameters as systematic uncertainties of the fit.
We assume the signal is relatively narrow, and fit the resonant signal with a Gaussian peak.
We do not attempt to perform a more complicated parametric fit to the background nor explore a different functional form. 
Instead, we assume that after the selection on the classifier, the mass distribution will fall smoothly,  and this will allow for a sufficient number of sideband events to reliably constrain the QCD multijet background.

Although we have not done so here, in a real implementation of an anomaly search one would want to use a cross-validation scheme so all the data could be used to search for a signal and none is ``wasted'' by the training (as in \cite{Collins:2018epr, Collins:2019jip}). 
This would involve splitting up the data into multiple samples and cycling through which samples are used for training and which were used for searching. 
Then one would simultaneously fit all of the signal regions to achieve full sensitivity. 

\subsection{Results}

\begin{figure}
    \centering
    \includegraphics[width = 0.49 \textwidth]{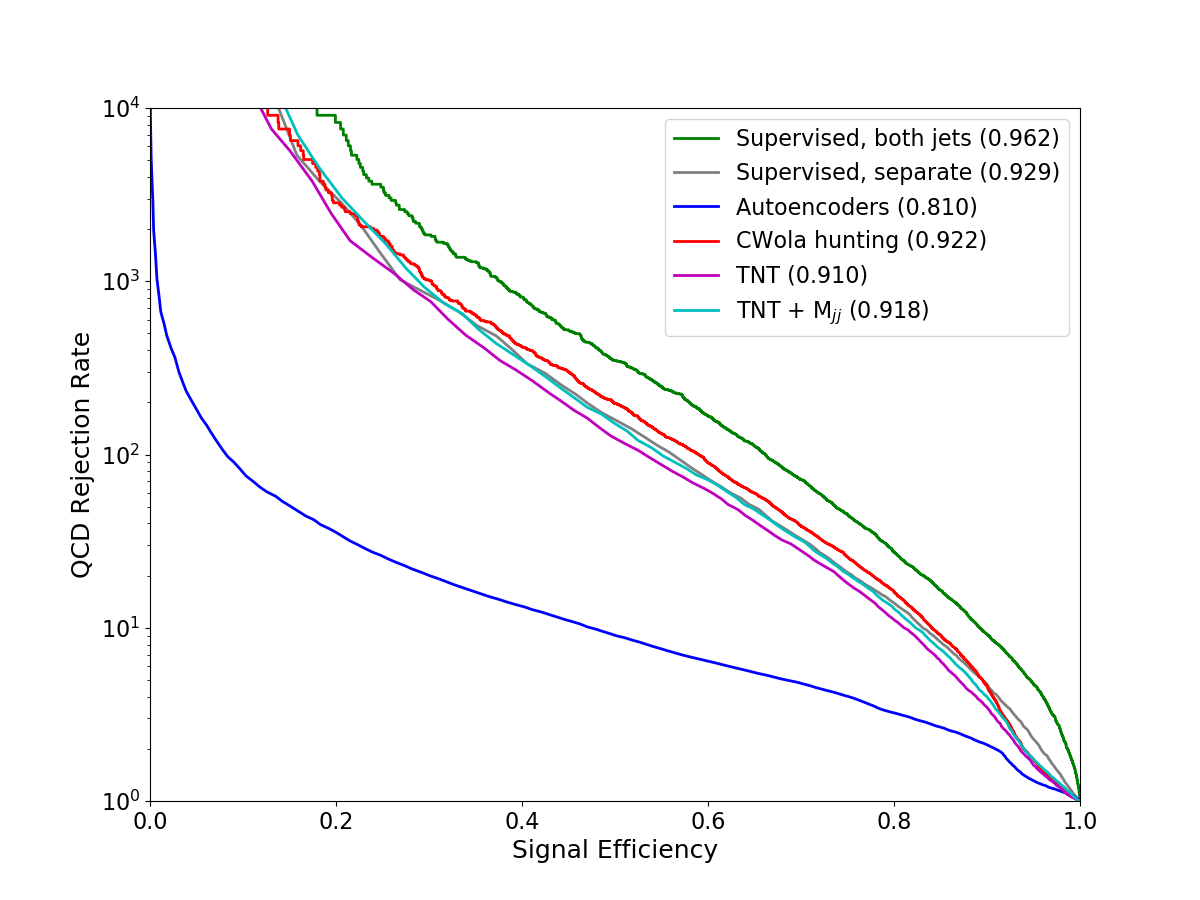}
    \includegraphics[width = 0.49 \textwidth]{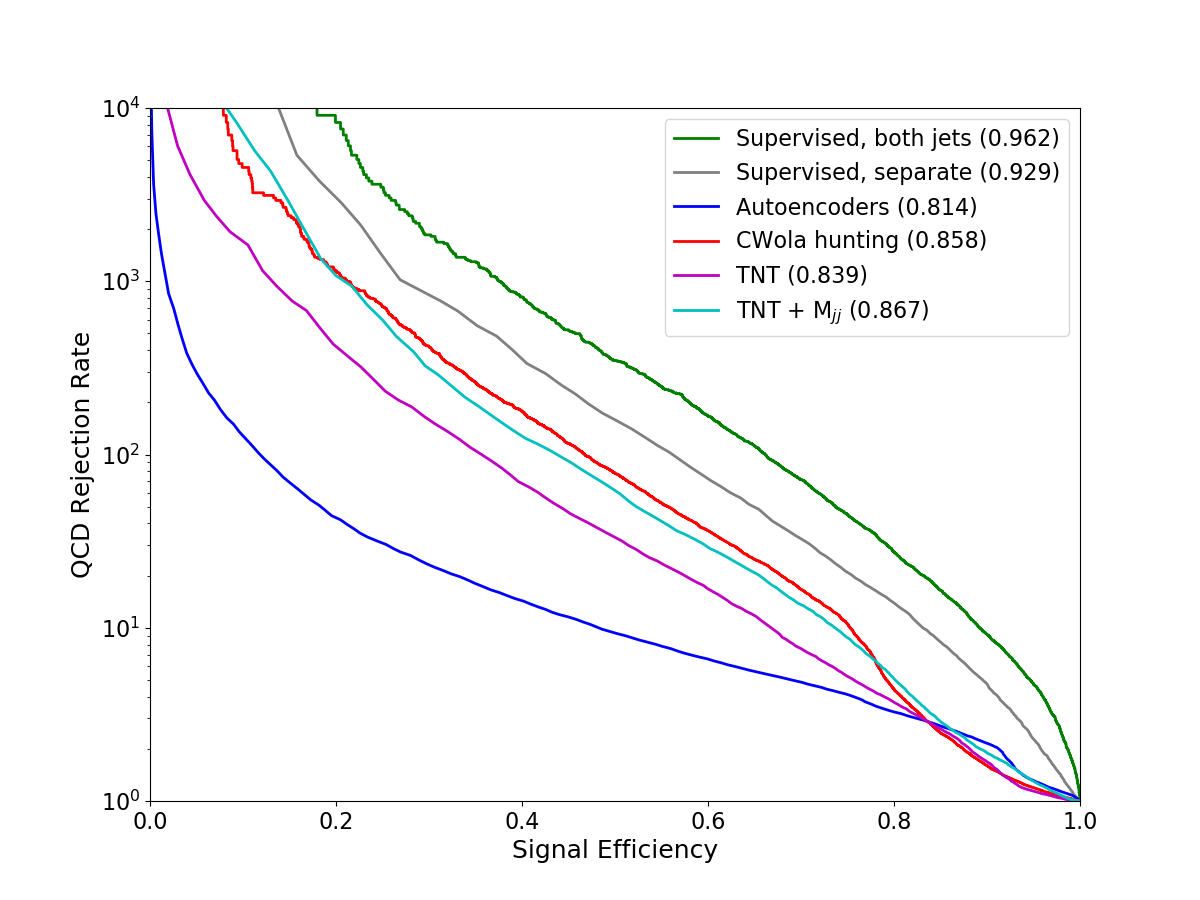}
    \includegraphics[width = 0.49 \textwidth]{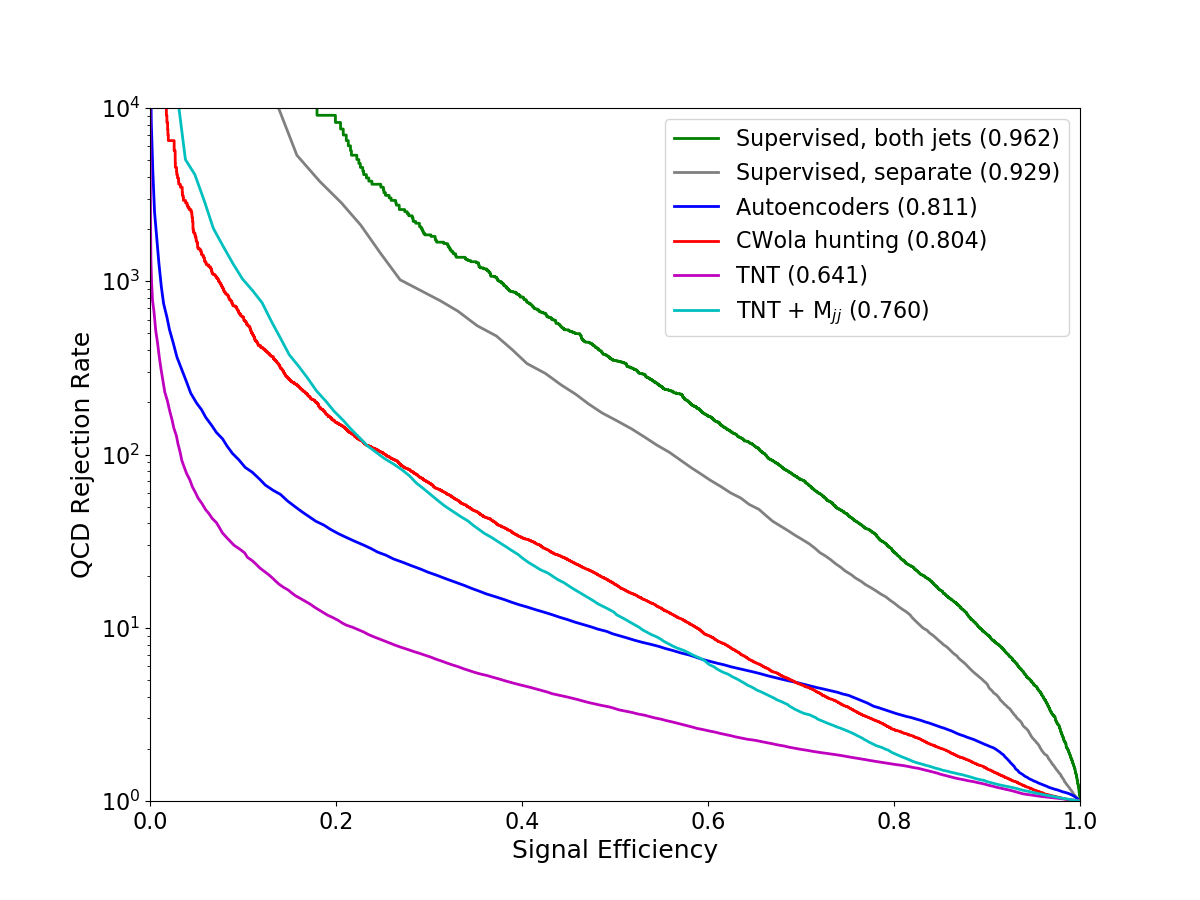}
    \includegraphics[width = 0.49 \textwidth]{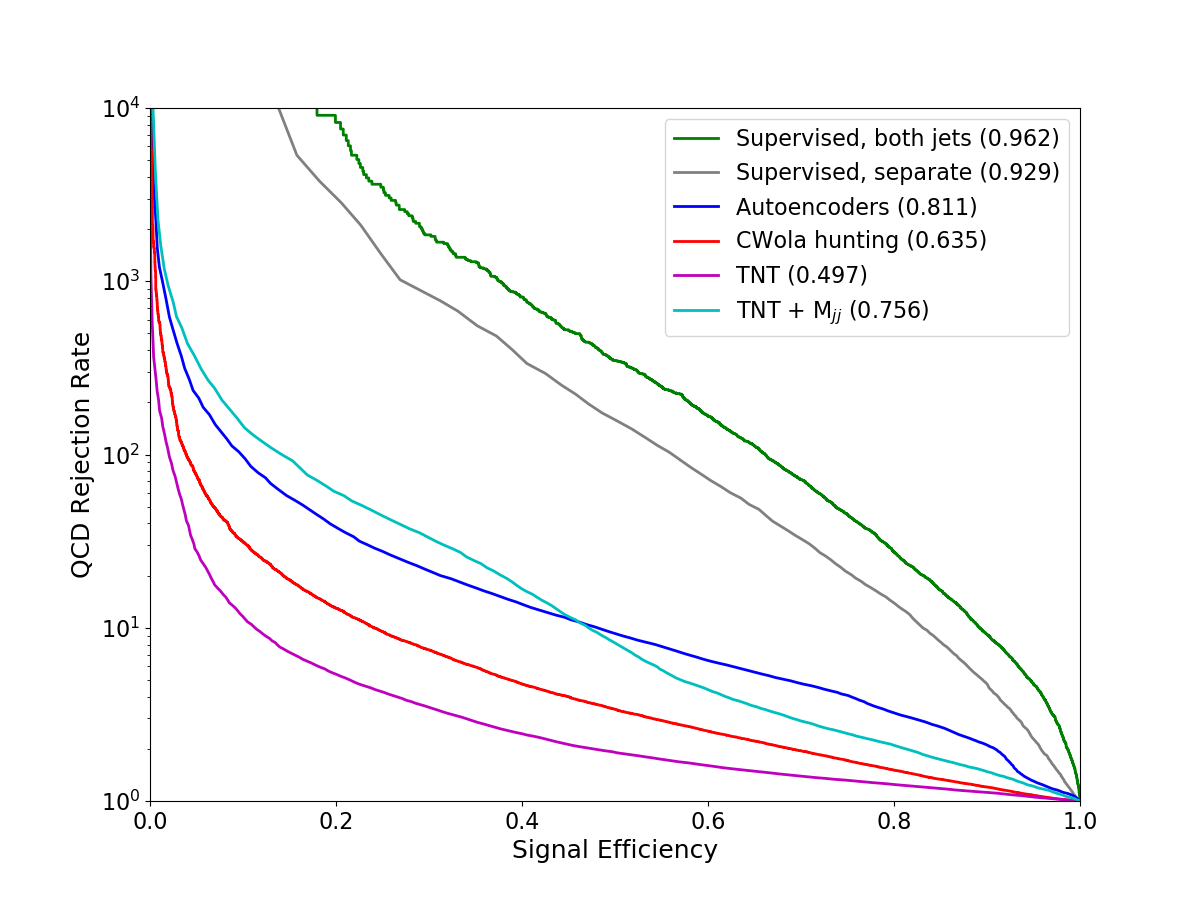}
    \caption{Classification performance for distinguishing signal and background events. 
    The numbers next to each label show the Area Under Curve (AUC) of each classifier.
    We show results for when the fraction of signal events is 9\%, 1\%, 0.3\% and 0.1\% in the top left, top right, bottom left and bottom right respectively. 
    We show the performance of the Tag N' Train technique with and without the dijet mass cut, and the CWoLa hunting and autoencoders.
    Two fully supervised classifiers trained with ground-truth labels, are shown for reference. 
    The one labeled 'separate' consists of 2 classifiers, trained on each jet separately and combined in the same way as the TNT classifiers, and the one labeled 'both jets' is trained on both jets at the same time.}
    \label{fig:roc_cmps}
\end{figure}

We evaluate the performance of the pure TNT classifiers, and the classifiers trained using TNT and a dijet mass cut (hereafter called TNT + M$_{jj}$ classifiers).
For this purpose, we use the sample of 300k events not used in the training step.
In Fig.~\ref{fig:roc_cmps}, we compare their performance against fully supervised classifiers, those trained using the CWoLa hunting method~\cite{Heimel:2018mkt, Farina:2018fyg} and autoencoders.
We also compare the performance of supervised classifiers trained using the images of both jets at the same time and classifiers trained on each jet individually and combined in the same way as the TNT classifiers.
One can see that there is a noticeable drop in performance when separating the jets, but that good classification performance is still possible. 
Although the 9\% signal test is rather optimistic from an anomaly search perspective, it shows that the both the TNT and TNT + M$_{jj}$ converge to the performance of a fully supervised classifier given sufficient signal.
For the 1\% signal test, the TNT classifier is somewhat worse than the TNT + M$_{jj}$ classifier, but still has significantly improved performance with respect to the autoencoder.
Finally, for the 0.3\% and 0.1\% signal tests, we can see that there is too little signal for the TNT classifier to learn from, and TNT performs significantly worse than the autoencoder. 
The TNT + M$_{jj}$ classifier performs similarly to CWoLa hunting for the 3 tests with larger signal, but for the 0.1\% test the TNT + M$_{jj}$ is able to maintain better performance better than the CWoLa hunting method, but without improving with respect to the autoencoders approach.

It is important to point out that because the signal chosen is a narrow resonance, it naturally favors the TNT + M$_{jj}$ and CWoLa hunting methods which assume this type of signal.
If the signal was a wide resonance, or a non-resonant signal, of similar cross section it is likely that these techniques will not perform as well, while the autoencoders and the regular TNT will have similar performance.
We leave a comparison of performance on non-resonant signals for future work. 

\begin{figure}
    \centering
    \includegraphics[width = 0.7 \textwidth]{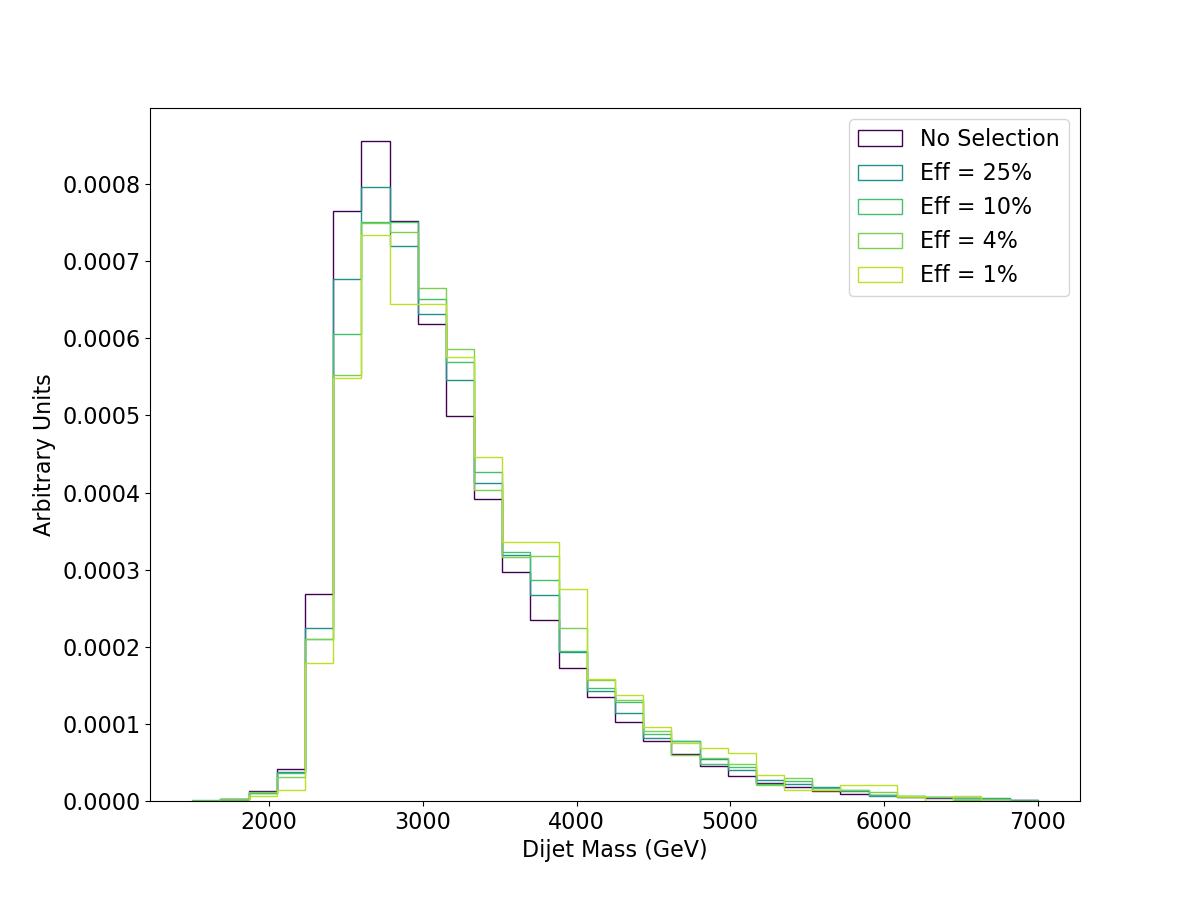}
    \caption{The QCD dijet mass distribution after applying different selections on the TNT + M$_{jj}$ classifiers, trained with the 1\% signal in the dataset. 
    The QCD dijet mass distribution remains smooth even down to a selection efficiency of 1\%, allowing the use of data-driven background estimates.}
    \label{fig:mass_sculpt}
\end{figure}

In addition to achieving good classification performance, we also highlight that neither the TNT or TNT + M$_{jj}$ classifiers  significantly sculpt the QCD dijet mass distribution.
In Figure~\ref{fig:mass_sculpt}, we show the QCD dijet mass distribution after applying various cuts using the TNT + M$_{jj}$ classifier. 
We can see that the shape of the distribution is not altered by any of these cuts. 
This is crucial because it allows the use of data driven estimates of the QCD background which rely on the smoothness of the dijet mass distribution. 
The lack of sculpting is due to our choice of classifier inputs, we normalize each jet image so that the sum of all pixel intensities is 1. 
This means that each image does not carry very much information about jet $p_T$ which can be used to sculpt the dijet mass distribution. 
But it is also important to point out algorithmic differences that can mitigate the risk of sculpting, between the TNT approach with a dijet mass cut and the CWoLa hunting. 
The first is that TNT selects the background-like and signal-like events using more information than just the dijet mass cut. 
This means that there are background events that populate the signal window and that are used in the training with a dijet mass in the signal-window, whereas in the CWoLa hunting approach all background events are in dijet mass sidebands. 
The other advantage is that by training a classifier for each jet separately, one can try to explicitly decorrelate the classifier's dependence on jet $p_T$ through one of the techniques that have been used in supervised jet classification \cite{Dolen:2016kst, Louppe:2016ylz, Shimmin:2017mfk, Bradshaw:2019ipy, Kasieczka:2020yyl}\footnote{In principle it is possible to do this in the CWoLa hunting approach as well if one trains a classifier for each jet separately, and attempts to decorrelate classification score from jet $p_T$.}.
We explored reweighting events in the background-rich sample to have the same $p_T$ distribution as the signal-rich region, but as there was not much mass sculpting to begin with, there were no significant differences in the mass sculpting or classification performance. 

\begin{figure}
    \centering
    \includegraphics[width = 0.48 \textwidth]{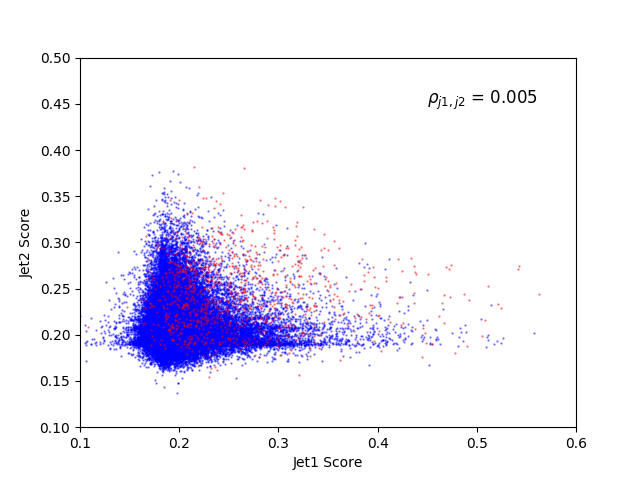}
    \includegraphics[width = 0.48 \textwidth]{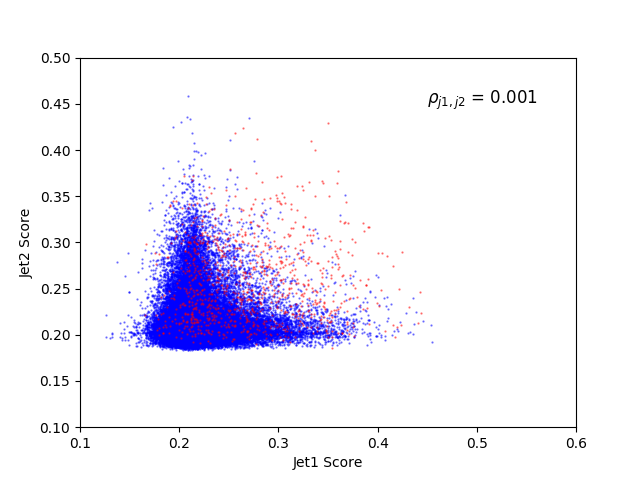}
    \caption{Correlation between classifier scores for the TNT classifier trained on a dataset with 1\% signal (left) and a classifier trained using CWoLa on mixed samples with similar signal to background ratio as the samples used in the TNT training (right).
    Blue dots are QCD background and red dots are signal.
    In the top right we report the Pearson correlation coefficient of the jet 1 and jet 2 scores for background events.}
    \label{fig:correlation}
\end{figure}

Another key feature to point out is that the TNT procedure maintains the independence of jet 1 and jet 2 scores on background events. 
In Fig.~\ref{fig:correlation} we show the correlation between the jet 1 and jet 2 classification scores for classifiers trained with TNT.
We also show a similar figure for `pure CWoLa' classifiers trained using randomly selected mixed samples with signal-rich and background-rich samples with similar S/B's to the TNT training.
We can see that the TNT classifier produces roughly similar distribution of jet 1 and jet 2 scores as the pure CWoLa classifier. 
We also compute the Pearson correlation coefficient between jet 1 and jet 2 scores for background events.
We can see that the TNT classifier does not develop a significant correlation between jet 1 and jet 2 scores. 
This is desirable from an event-level classification standpoint because it means when a background event passes the selection for one of the classifiers it is not biased to be more likely to pass the other.
Additionally this independence allows the use of background estimation techniques that rely on creating control regions by inverting the selection on one of the jets.

We also test the signal extraction procedures detailed in the last section using this sample of 300k events.

\begin{figure}
    \centering
    \includegraphics[width = 0.75 \textwidth]{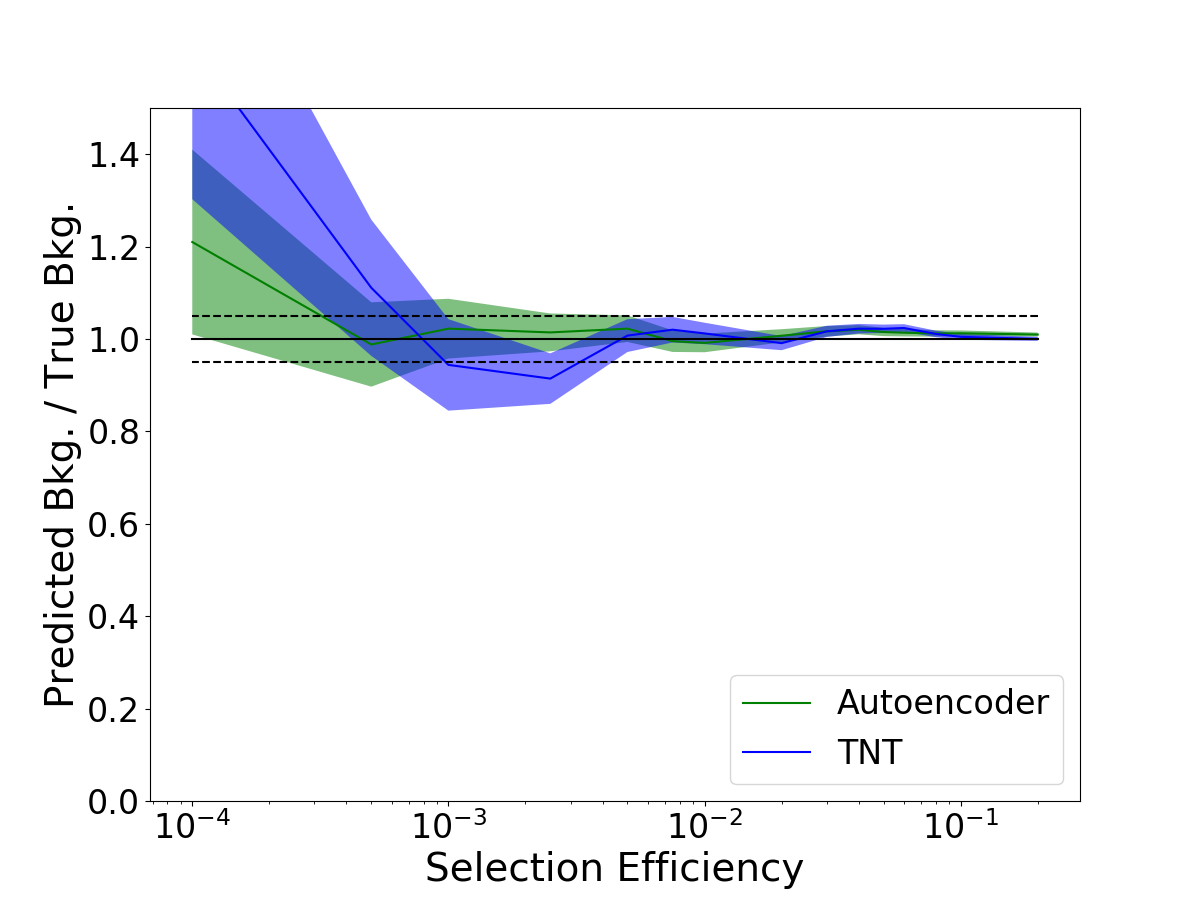}    
    \caption{A comparison of the estimated and true number of background events as a function of selection efficiency for the cut and count based analysis. 
    The shaded regions around the central values show the 1$\sigma$ uncertainties due to the Poisson uncertainty on the observed number of background events. The black dashed lines shows a 5\% around the ratio of 1. }
    \label{fig:bkg_check}
\end{figure}

To test the cut-and-count approach, we use the sample with 1\% signal and the TNT classifiers and autoencoders to select events. 
In Figure \ref{fig:bkg_check} we compare our estimated number of background events to the true number of events at various selection efficiencies.
One can see that this method of background estimation works quite well even down to selection efficiencies as small as $10^{-3}$ for both the autoencoders and the TNT classifiers. 

We also do a rough comparison of the expected and observed significance at different selection efficiencies based on this background estimate. We take the uncertainty on the background estimate to be the combination of the statistical uncertainty (from the selection efficiency measurements) and a 3\% systematic uncertainty. We compute the significance as: \\
\begin{equation}
\centering
\sigma_{obs} = \frac{N_{obs} - N_{pred}}{\sqrt{N_{obs} + \sigma_{b}^2}} \quad \quad            
\sigma_{exp} = \frac{N_{sig}}{\sqrt{N_{obs} + \sigma_{b}^2}},
\end{equation}
where $N_{obs}$ is the number of events observed after selection, $N_{pred}$ is the number of predicted background events, $\sigma_{b}$ is the total uncertainty on the background estimate, and $N_{sig}$ is the true number of signal events after selection.  

We compare the significance of the autoencoder and TNT based approaches in Table \ref{tab:significance}. 
When using the autoencoders, the optimal working point achieves a $~3\sigma$ significance while the TNT classifier can achieve a significance $>10\sigma$. Of course such a high value of significance should not be taken literally, especially for this rough estimate, but this does illustrate the potential gains of the TNT approach. 

\begin{table}[!htbp]
    \centering
         \caption{A rough comparison of the expected and observed significance (in $\sigma$'s) for the autoencoder and Tag N' Train based searches on a dataset with 1\% signal. The shown significance is based on the total uncertainty on the background estimate and the Poisson uncertainty on the number of observed events. }
     \label{tab:significance}
    
    \resizebox{\textwidth}{!}{\begin{tabular}{|c |c | c | c | c|}

    \hline
    Selection Efficiency & Sig. Exp. (AE) & Sig. Obs. (AE) & Sig. Exp. (TNT) & Sig. Obs. (TNT) \\
    \hline
    10\% & 1.8 & 1.4 & 2.3 & 2.1 \\
    5\% & 2.1 & 1.7 & 3.9 & 3.2 \\
    3\% & 2.5 & 2.0 & 5.6 & 5.1 \\
    1\% & 2.9 & 3.1 & $>$ 10 & $>$ 10 \\
    0.5\% & 2.8 & 2.3 & $>$ 10 & $>$ 10  \\
    \hline

    \end{tabular}}
\end{table}

For the bump hunt approach, we try samples with 0.3\% and 0.1\% signal and select events with the TNT + $M_{jj}$, CwoLa hunting, autoencoder and supervised classifiers.
For each classifier we select events with an overall efficiency of 3\%. 
In the presence of a signal a dijet mass peak forms at the signal hypothesis, after a percentile selection on our data. 
We fit for the presence of a resonant signal modeled as a Gaussian shape peaking at 3.5~TeV.
We verified that when no signal is present in our sample, no significant bump is created by our procedure.
In Tables~\ref{tab:sigfit03} and \ref{tab:sigfit01}, we compare the significance on samples with 0.3\% and 0.1\% signal respectively.  
We compare the significance obtained prior to any selection and after events have been selected using each classifier. 
In both cases we observe that the significance prior to any selection is $< 2 \sigma$ meaning the signal would have gone unnoticed in 
an inclusive dijet resonance search on a dataset of this size. 
We observe that for the 0.3\% case the signal significance roughly follows the classification performance shown in Fig.~\ref{fig:roc_cmps},
with both the $TNT + M_{jj}$ and CWoLa hunting classifiers obtaining significances close to $5\sigma$.
For the 0.1\% case we see a large drop off in significance enhancement for all three of the anomaly classifiers, 
with all of the significances $< 2 \sigma$ while the supervised classifier can still clearly find the signal.

\begin{table}[!htbp]
    \centering
    \caption{A comparison of the observed significance (in $\sigma$'s) of a bump hunt search using various different classifiers to select events with a 3\% efficiency.
      We also compare to the significance of a bump hunt on events prior to any classifier selection.
     Except for the supervised, the classifiers were trained on a sample with 0.3\% of signal events. All significances are obtained on a validation sample that also has 0.3\% signal events.}
     \label{tab:sigfit03}
    
    \resizebox{\textwidth}{!}{\begin{tabular}{|c c  c  c  c|}

    \hline
    No Selection &  AutoEncoder & CWOLA Hunting & TNT+ M$_{jj}$ & Supervised \\
    \hline
    1.4 & 3.0 & 4.8 & 5.3 & $>$10 \\
    \hline

    \end{tabular}}
\end{table}

\begin{table}[!htbp]
    \centering
    \caption{A comparison of the observed significance (in $\sigma$'s) of a bump hunt search using various different classifiers to select events with a 3\% efficiency.
      We also compare to the significance of a bump hunt on events prior to any classifier selection.
    Except for the supervised, the classifiers were trained on a sample with 0.1\% of signal events. All significances are obtained on a validation sample that also has 0.1\% signal events.}
     \label{tab:sigfit01}
    
    \resizebox{\textwidth}{!}{\begin{tabular}{|c c  c  c  c|}

    \hline
    No Selection & AutoEncoder & CWOLA Hunting & TNT+ M$_{jj}$ & Supervised \\
    \hline
    0.9& 0.9& 1.4 & 1.4 & 8 \\
    \hline

    \end{tabular}}
\end{table}

We remark that the p-value computed in the bump-hunt approach is only a local p-value. 
The translation to a global p-value would depend on the procedure used to scan over the full dijet mass range and how many different selections are tried in each mass window \footnote{Because the p-values observed with different selections in the same mass window will be correlated, computation of a global p-value is non-trivial. We leave the exploration of this for future work.}. 

\begin{figure}
    \centering
    
    \includegraphics[width = 0.45 \textwidth]{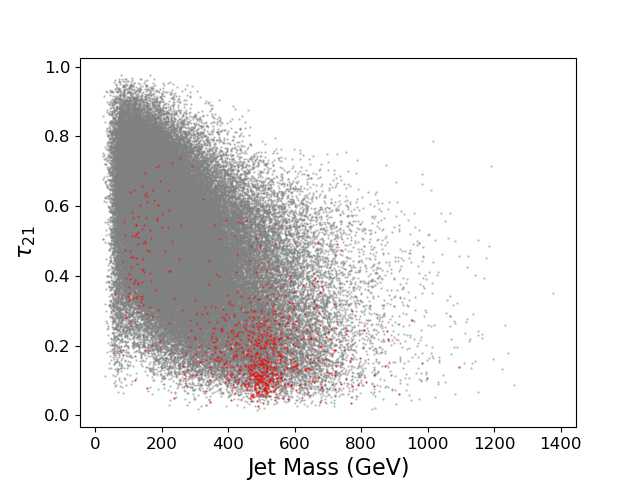}
    \includegraphics[width = 0.45 \textwidth]{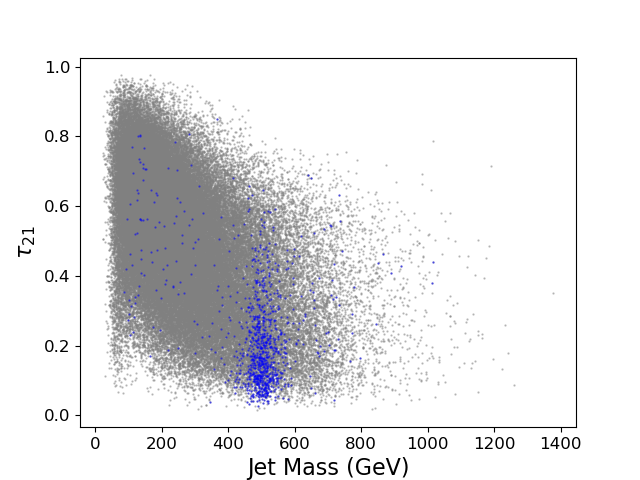}
    \includegraphics[width = 0.45 \textwidth]{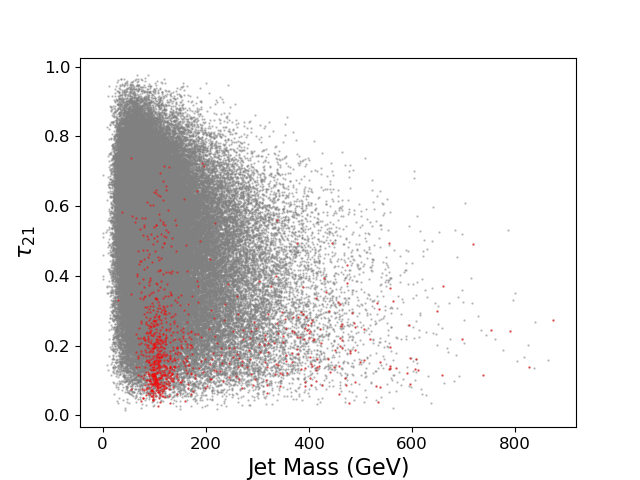}
    \includegraphics[width = 0.45 \textwidth]{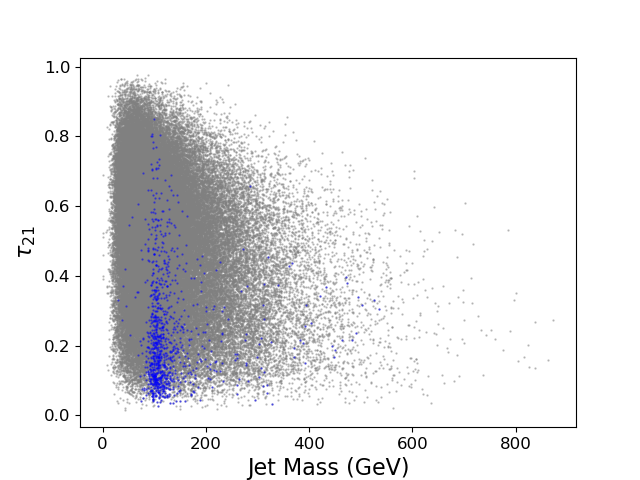}

    \caption{Correlation between $\tau_{21}$ and jet mass for the dataset with 1\% signal. The top row corresponds to the heavier jet and the bottom row the lighter jet. 
    On the left (in red) are the 1\% of events the TNT classifier found to be most signal-like and on the right (in blue) the true signal events.}
    \label{fig:signal}
\end{figure}

If a local p-value is below some threshold, it would be crucial to characterize the nature of the signal. 
While there are some known strategies that can be used to understand what a deep CNN has learned \cite{kim2017interpretability,simonyan2013deep, olah2017feature}, a simple approach would be to just examine the events that the classifier has found to be most signal like.
In Figure \ref{fig:signal} we compare the characteristics of the events the TNT classifier found to be most signal-like to the characteristics of the true signal events. 
We show 2D scatter plots of jet mass and the N-subjetiness ratio $\tau_{21}$ \cite{Thaler:2010tr}.
Despite not using the jet mass or N-subjetiness as direct inputs to the network, one can see that the TNT classifier has learned the correct masses of the X and Y jets and that they are both two pronged.
Characterizing the signal in this way would also give an analyzer confidence they had truly found evidence of new physics rather than a unknown feature of the detector. 

\section{Conclusions}
\label{sec:conclude}

We have introduced a new method of training classifiers directly on data called Tag N' Train. 
It relies on decomposing the data into two distinct sub-objects which can be classified separately.
One can then use one of the sub-objects to tag events as signal-like or background-like, and those samples can be used to train a classifier for the other object.
Here we have explored the possibility of using the Tag N' Train technique to perform a dijet anomaly search by using autoencoders trained directly on data as the initial classifiers. 
We demonstrate that given sufficient signal in the data, the TNT technique is able to produce classifiers that perform significantly better than the autoencoders. 
When a cut on the dijet mass is used in addition to the autoencoders to select signal-like events, the TNT classifiers achieve similar performance to those trained using the CWoLa hunting technique. 

As this work was meant to be a proof of concept for the Tag N' Train method, we believe there is substantial room to improve the performance of the TNT dijet anomaly search, both by optimizing the initial network used to detect anomalous jets, and the architecture of the classifier trained with TNT.
An obvious direction to explore would be other variants of the autoencoder architecture, such as variational autoencoders~\cite{kingma2013autoencoding, Cerri:2018anq} or normalizing-flow based autoencoders~\cite{kingma2016improving}, but in principle any anomaly detection method that is able to isolate signal events using only one jet at a time could work as an initial classifier.
Also, while using low-level inputs like jet images to the TNT classifiers offers robustness to many types of signals using higher-level features may offer advantages as well.
If there was only a small amount of signal present, it would likely be easier for the network to learn if higher level features were used.
However, by restricting the information given to a hand-selected subset of variables, one may lose sensitivity to anomalies exhibiting novel features. 
We leave the exploration of these ideas for future work. 

Although here we have applied the TNT technique to a dijet resonance search, the performance of the TNT technique without using dijet mass window, shows it could be applied to a non-resonant anomaly search as well.
We have also demonstrated that Tag N' Train technique naturally pairs with an ABCD background estimate due to the two classification scores of the two objects are independent by design.

A key point to explore for the future would be how the Tag N' Train technique performs in the presence of sub-dominant SM backgrounds with interesting sub-structure, such as top quark pairs or W+Jets production.
Preventing the TNT technique from learning these events as signal-like may require additional control regions to be added to the background-rich sample in training, or explicitly veto-ing events that look like from known SM backgrounds from the signal-rich sample during training. 

The Tag N' Train framework could also be applied to model-specific search as well. 
Running the Tag N' Train technique with classifiers for Standard Model jets as inputs, while scanning for a resonance, could target models covered by existing searches.
The possible advantage of the Tag N' Train framework would be that by training new classifiers directly on data, one would mitigate the effects of imperfections in the simulation used in training. 
It would be interesting to compare how the performance of this sort of Tag N' Train search compares to a traditional supervised search if there were significant mis-modeling of the signal or background in simulation.

\section*{Code and data availability}

Code to reproduce all of our results can be found on \href{https://github.com/OzAmram/TagNTrain}{Github} and the dataset used is available on Zenodo~\cite{Kasieczka_LHCO_RD}.

\acknowledgments

We thank Petar Maksimovic for useful input and his continuous support. 
We also would like to thank the organizers of the ML4Jets conference, and in particular Gregor Kasieczka, Ben Nachman, and David Shih for organizing the LHC Olympics 2020 challenge that was the impetus for this work.

\bibliographystyle{jhep}
\bibliography{main}

\appendix
\section{Attempting to Train Using a Single Object}
\label{sec:same_jet_test}

\begin{figure}
    \centering
    \includegraphics[width = 0.8 \textwidth]{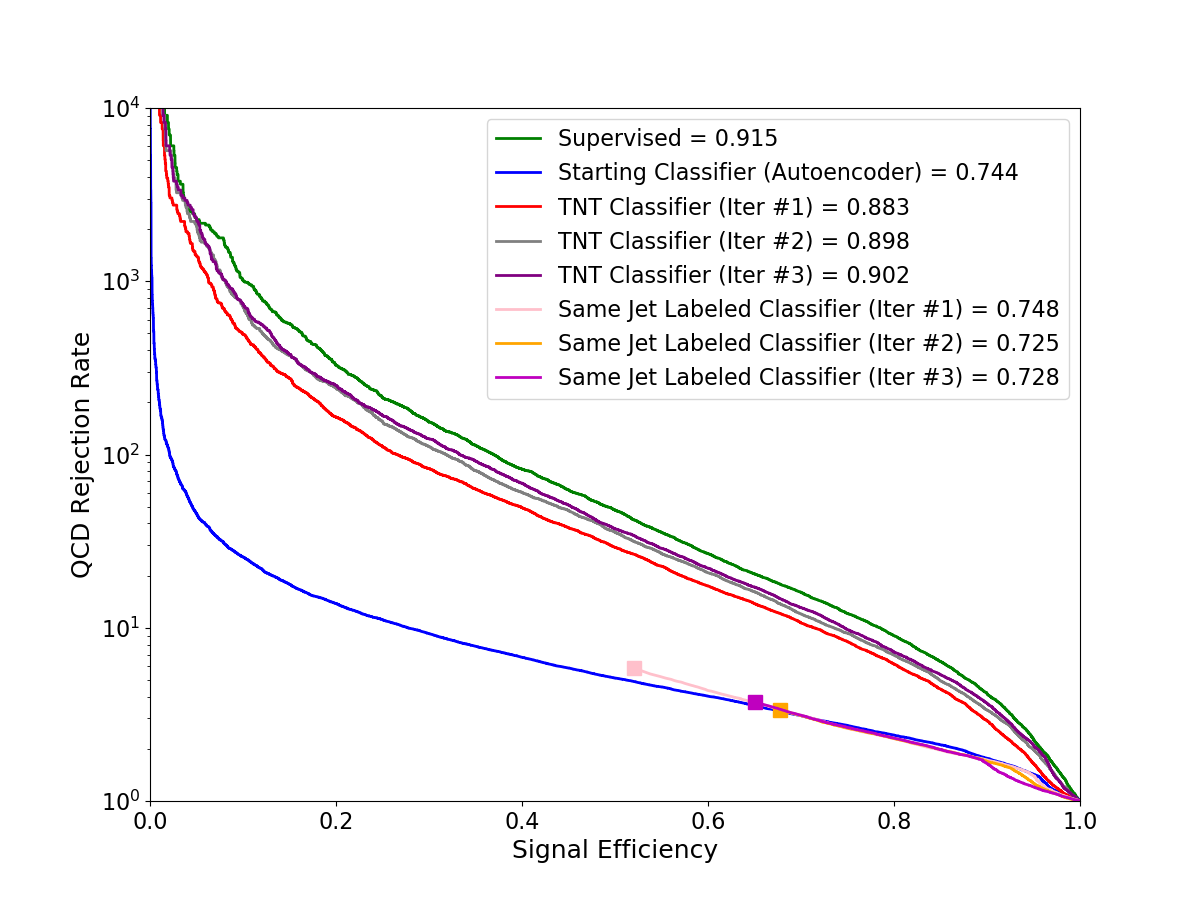}
    \caption{A comparison of classifiers trained using the same jet repeated versus the TNT approach. The ROC curves of the same-jet-trained classifiers are truncated because selections tighter than this point are not possible due to the classifier scores of these events all being exactly the same.}
    \label{fig:roc_samejet}
\end{figure}

To demonstrate why the Tag N' Train approach is necessary, we show here that one cannot use a similar approach but with only one object to train improved classifiers on unlabeled events.
Using the same dataset and setup as in the main text, we attempt to train a new classifier for jet-1 using anomalous events selected based on their jet-1 anomaly scores (rather than jet-2 scores as in the TNT approach). 
To test whether this is possible at all, we use a dataset with a very large amount of signal, 9\%. 
Starting with an autoencoder as an initial classifier, we select the 20\% of events with the highest classifier scores to be the signal-rich sample and the 40\% of events with the lowest scores to be the background-rich sample. 
We then train a classifier to distinguish between these two samples.
Similar to the TNT approach, this new classifier is then used with a new set of data events to train another classifier.
We repeat for a total of 3 iterations, with 200,000 events being used in each iteration.
In Figure \ref{fig:roc_samejet} we compare the performance of classifiers trained with this method to one trained with TNT.

One can see that the classifier trained repeatedly with the same jet is not able to perform any better than the autoencoders used for initial classification even after 3 iterations, while the TNT classifier greatly surpasses the autoencoder in the first iteration and improves further in the second iteration. 
Furthermore, we observe that the classification scores for the same-jet classifiers all plateau to the exact same value, making selections with low efficiency impossible. 
This behavior can be understood by considering the training procedure.
First of all, the same-jet classifier is not able to learn any more information about the signal than the autoencoders because its training objective was to classify the jets in the exactly the same way as the labeling classifier.
Additionally, the collapse of classification scores to a single value occurs because the labeling procedure 'collapses' the signal-like events to all have a labels of 1.
This issue could likely be solved by using a soft-labeling scheme, but the fundamental challenge of surpassing the performance of the initial classifier remains. 
The Tag N' Train approach avoids these issues by using information external to the classification task (namely the other object in the event) to construct the signal- and background-rich samples.

\end{document}